\documentclass[a4paper,12pt]{article}
\usepackage{graphicx}
\usepackage{mathptmx}
\usepackage{amsmath}

\begin{document}
\vspace*{3\baselineskip}

\hangindent=0.55\textwidth \hangafter=0
\noindent
{\small binding neuron, biological and medical cybernetics, interspike interval distribution,  complex systems, cognition and systems}
\vskip 3\baselineskip

\noindent
{\large Alexander VIDYBIDA}
\vskip 3\baselineskip

\begin{center}
{\Large\bf OUTPUT STREAM OF BINDING NEURON\\[8pt] WITH DELAYED FEEDBACK}
\vskip 3\baselineskip
\end{center}

\hangindent=\parindent\hangafter=0\noindent
{\small  A binding neuron (BN) whith delayed feedback is considered. The neuron is fed externally with a Poisson stream of intensity $\lambda$. The neuron's output spikes are fed into its input with time delay $\Delta$. The resulting output stream of the BN is not Poissonian, and we look for its interspike intervals (ISI) distribution. For BN with threshold 2 an exact mathematical expression as function of $\lambda$, $\Delta$ and BN's internal memory, $\tau$ is derived for the ISI distribution, and for higher thresholds it is found numerically. The distributions found are characterized with discontinuities of jump type, and include singularity of Dirac's $\delta$-function type.
 It is concluded that delayed feedback presence can radically  alter neuronal output firing statistics.}
\vskip 2\baselineskip

\section{Introduction}

The role of input spikes timing in functioning of either single neuron, or neural net has been addressed many times, as it constitutes one of the main problem in neural coding. The role of timing was observed in processes of perception \cite{Laurent}, memory \cite{Hebb}, objects binding and/or segmentation \cite{Eckhorn}. At the same time, where does the timing come from initially? 
In reality, some timing can be inherited from the external world during primary sensory reception. In auditory system, this happens for the evident reason that the physical signal, the air pressure time course, itself has pronounced temporal structure in the millisecond time scale, which is retained to a great extent in the inner hair cells output \cite{Cariani,Eggermont,Aud}.  In olfaction, the physical signal is produced by means of  adsorption-desorption of odor molecules, which is driven by Brownian motion.
In this case, the primary sensory signal can be represented as Poisson stream, thus has not any remarkable temporal structure.

 Nevertheless, temporal structure can appear in the output of a neuron fed by a structureless signal. 
After primary reception, the output of corresponding receptor cells is further processed in primary sensory pathways, and then in higher brain areas. During this processing, statistics of poststimulus spiking activity undergoes substantial transformations (see, e.g. \cite{Eggermont}). After these transformations, the eventual activity is far away from the initial one. This process is closely related to the information condensation \cite{Konig}.

We now put a question: What kind of physical mechanisms might underlie these transformations? It seems that, among others, the following features are responsible for spiking statistics of a neuron in a network: (i) several input spikes are necessary for a neuron from a higher brain area to fire an output spike (see, e.g. \cite{Andersen}); (ii) a neural net has numerous interconnections, which bring about feedback and reverberating dynamics in the net. Due to (i) a neuron must integrate over a time interval in order to gather enough input impulses to fire. As a result, in contrast to Poisson stream, the shortest ISIs between output spikes will no longer be the most probable. This was observed long ago \cite{Segundo} in numerical experiments with leaky integrate-and-fire (LIF) neuronal model and confirmed recently in exact mathematical derivation for binding neuron \cite{Vid4}. Due to reverberation, an individual neuron's output impulses can have some delayed influence on the input of that same neuron. This can be the source of positive feedback which results in establishing of dynamics partially independent of the stimulating input (compare with \cite{Konig}), and which governs neuronal spiking statistics.

In this paper, we consider a simplest possibility to test influence of (i), (ii), above on neuronal firing statistics. As neuronal model we take the binding neuron (BN) one. Exact mathematical expression is derived for output ISI distribution as a function of input Poisson stream intensity, $\lambda$, BN's internal memory, $\tau$, delay value in the feedback line, $\Delta$, when BN has threshold 2. For higher thresholds the distributions are calculated numerically, by means of Monte Carlo algorithm. The distributions found are characterized with discontinuities of jump type, and include singularity of Dirac $\delta$-function type. It is concluded that delayed feedback presence can radically  alter neuronal output firing statistics.

\begin{figure} [h]
\unitlength=0.9mm
\begin{center}
\includegraphics[width=0.75\textwidth]{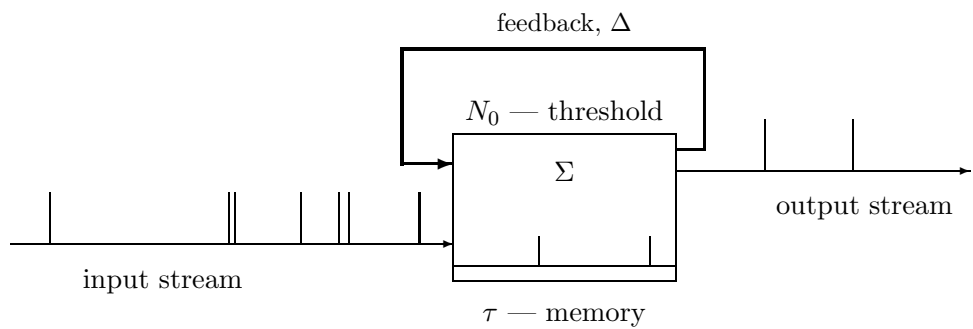}
\end{center}
\caption{\label{BNF} Binding neuron with feedback (see \cite{Vid3} for deatils). 
$\tau$ is similar to the ``tolerance interval'' discussed in \cite[p. 42]{MacKay}. Multiple input lines with Poisson streams are joined into a single one here.}
\end{figure}

\section{Methods}

For analytical calculations we consider threshold value $N_0=2$.

\subsection{BN without feedback}

The binding neuron model \cite{Vid3} is inspired by numerical simulation of Hodg\-kin-Hux\-ley-type point neuron \cite{Vid}, as well as by the leaky in\-teg\-rate-and-fire (LIF) model \cite{Segundo}. In the binding neuron, the trace of an input is remembered for a fixed period of time after which it disappears completely. This is in the contrast with the above two models, where the postsynaptic potentials decay exponentially and can be forgotten only after triggering. The finiteness of memory in the binding neuron allows one to construct fast recurrent networks for computer modeling as well as obtain exact mathematical conclusions concerning firing statistics of BN. Recently, \cite{Vid4}, the finiteness is utilized for exact mathematical description of the output sto\-c\-h\-astic process if the binding neuron is driven with the Poisson input stream. 

The BN works as follows (see Fig. \ref{BNF} with the delay line removed). Each input impuls is stored in the BN for a fixed period of time, $\tau$, and then is forgotten. When the number of stored impulses, $\Sigma$, becames equal or bigger then the BN's threshold, $N_0$, the BN fires output spike, clears its internal memory, and is ready to receive fresh inputs. Normally, any neuron has a number of input lines. If input stream in each line is Poisson, all lines can be joined into a single one, like in Fig. \ref{BNF}, with intensity, $\lambda$, equal to sum of intensities in the individual lines. 

Recently \cite{Vid4}, the output statistics was calculated for this model with $N_0=2$. In this work we will need the following exact expressions from \cite{Vid4}.
ISI distribution probability density function, $P^0(t)$, where $t$ denotes the output ISI duration,
\begin{multline}\label{P0}
m\tau\le t\le (m+1)\tau\quad \Rightarrow\quad
P^0(t)\,dt
=e^{-\lambda t}\frac{\lambda^{(m+1)}(t-m\tau)^{(m+1)}}{(m+1)!}\,\lambda\, dt
+
\\
+e^{-\lambda t}\sum\limits_{1\le k\le m}\frac{\lambda^k}{k!}
\left((t-(k-1)\tau)^k
-(t-k\tau)^k\right)\,\lambda\, dt,\quad m=0,1,\dots ,
\end{multline}
and the first moments, $W_1$ of the distribution (\ref{P0}),
\begin{equation}\label{W1}
W_1\equiv\int\limits_0^\infty t\,P^0(t)\,dt=
\frac{1}{\lambda}\left(2+{1\over e^{\lambda\tau}-1}\right),
\end{equation}

\subsection{Fixed feedback line}

Any output impuls of BN with feedback line (BNF) may be produced either with impulse from the line involved, or not. 
We assume that, just after firing and sending output impulse, the line is never empty. This assumption is selfevident for output imulses produced without impulse from the line, or if the impulse from the line was involved, but entered empty neuron. In the letter case, the second (triggering) impulse comes from the Poisson stream, neuron fires and output impuls goes out as well as enters the empty line. On the other hand, if impulse from the line triggers BN, which already keeps one impulse from the input stream, it may be questionable if the output impulse is able to enter the line, which was just filled with the impulse. We expect it does. This means biologically that we ignore the refraction time - a short period necessary for a nervous fibre to recover from conducting previous spike before it is able to serve for the next one. Thus, at the beginning of any output ISI, the line keeps impulse with time to live $s$, where $s\in]0;\Delta]$. 

It is clear, that variability of the input Poisson stream should be combined with the variability in $s$ value in order to calculate the output stream properties, like ISI probability density $P^\Delta(t)$. In this  subsection, we define an auxiliary probability density, $P^\Delta_s(t)$,
in which the $s$ is put fixed at the beginning of any output ISI. Thus, instead of considering a stationary firing process in which both firing moments and $s$ are determined by the input Poisson process, we consider a process in which, after each firing, the line keeps impulse with time to live equal $s\in]0;\Delta]$.

\begin{figure}
\begin{center}
\includegraphics[width=0.66\textwidth,angle=0]{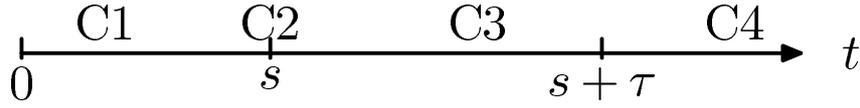}
\end{center}
\caption{\label{ts-cases}%
Domains of $t$ used for calculating $P^\Delta_s(t)$.}
\end{figure}

In order to derive $P^\Delta_s(t)$ it is suitable to separate possible values $t$ of ISI duration into several groupes as shown in Fig. \ref{ts-cases}.

In case C1, $t<s$. Here output impulse must be triggered without the line impulse involved. Therefore, distributions for such ISI values is the same as for BN without feedback:
\begin{equation}\label{vyp1}
P^\Delta_s(t)=P^0(t),\quad t<s.
\end{equation}

Consider case C2. The probability to obtain ISI exactly equal to $s$ is not infinitesimally small. This event is equivalent to the event $A_{\text{S}_1}(s)$ that BN starts empty at moment 0 and appears without triggerings in state S$_1$ (keeps impulse) at moment $s$. In order to obtain the probability $P\{A_{\text{S}_1}(s)\}$, let us take into account that $P^0(s)\,ds$ can be obtained as the product of $P\{A_{\text{S}_1}(s)\}$ and the probability to get input impulse in infinitisemal interval $ds$, which is $\lambda\,ds$. Therefore,
\begin{equation}\label{PAC1}
P\{A_{\text{S}_1}(s)\}
= \frac{P^0(s)}{\lambda},
\end{equation}
which together with Eq. (\ref{P0}) gives the $\delta$-function's mass in the expression for $P^\Delta_s(t)$ at point $t=s$.

In order to keep expressions shorter, let us assume that $\Delta<\tau$, and calculate ISI distribution for the case C3, above. Due to the assumption made, the probability to obtain ISI value $s<t \le s+\tau$ is just equal to the probability that first input impuls comes at required moment $t$. Therefore,
\begin{equation}\label{gustdt}
P^\Delta_s(t)=e^{-\lambda t}\lambda,\quad  s<t\le s+\tau.
\end{equation}

Consider case C4,\, $t\ge s+\tau$. It is realised if three independent events occure in series: (i) $A_{\text{S}_0}(s)$; (ii) interval $]s;s+\tau[$ is free from input impulses; (iii) BN without feedback starts from state S$_0$ at moment $s+\tau$ and is firstly triggered at moment $t$. These events are independent since their realizations are defined by behavior of Poisson input stream on intervals, which are mutually disjoint.  Due to the assumption made, the probability to have both (i) and (ii) is the same as to have in the Poisson input stream an ISI longer then $s+\tau$, and (iii) has the probability $P^0(t-s-\tau)\,dt$. Thus,
\begin{equation}\label{rozpdt}
P^\Delta_s(t)=
e^{-\lambda (\tau+s)}
P^0(t-s-\tau)\quad t\ge s+\tau\,.
\end{equation}

Equations (\ref{vyp1}), (\ref{PAC1}), (\ref{gustdt}), (\ref{rozpdt}) can be written together as sum of singular and regular parts:
\begin{equation}\label{rozpsum}
P^\Delta_s(t)=P^{\Delta r}_s(t)+P^{\Delta s}_s(t),
\end{equation}
where
\begin{equation}\label{rozpds0}
P^{\Delta s}_s(t)=e^{-\lambda s}\lambda s\,\delta(t-s),
\end{equation}
\begin{equation}\label{rozpdr0}
P^{\Delta r}_s(t)=
\begin{cases}
e^{-\lambda t}t\lambda^2,\quad t\in]0;s],\hfill(**)\\\\
\lambda\,e^{-\lambda t},\quad  t\in]s;s+\tau],\hfill(*)\\\\
e^{-\lambda (\tau+s)}P^0(t-s-\tau),\quad s+\tau\le t\quad \hfill(*)
\end{cases}\,,
\end{equation}
where assumption $\Delta<\tau$ is taken into account.

\subsection{Derivation outline}

When initial data is forgotten, the firing process of BN with delayed feedback becomes stationary. This brings about a stationary distribution, $f(s)$, for time to live, $s\in]0;\Delta]$, of an impuls in the feedback line at the moment of beginning of any output ISI. Having exact expression for $f(s)$, one could calculate required output ISI distribution as follows:
\begin{equation}\label{usered}
P^\Delta(t)=\int\limits_0^\Delta
P_s^\Delta(t) f(s)\,ds.
\end{equation}

In order to find $f(s)$, consider the transition probabilities $P(s\mid s'),\quad s,s'\in ]0;\Delta]$, which give probability that at the begining of some output ISI, the line has impulse with time to live $s$, provided that at the beginning of the previous ISI it had impulse with time to live $s'$. $P(s\mid s')$ can be found based on known expression for $P_s^\Delta(t)$. $f(s)$ is then found as normed to 1 solution to the following equation:
\begin{equation}\label{umoriv}
\int\limits_0^\Delta P(s\mid s')\,f(s')\,ds'=f(s).
\end{equation}

\section{Main calculation}
\subsection{Transition probabilities}

From the meaning of $P_s^\Delta(t)$ it follows that Eq. (\ref{rozpdr0})($**$) allows to calculate $P(s\mid s')$ for $s<s'$:
\begin{equation}\label{umov0}
P(s\mid s')=e^{-\lambda(s'-s)}\lambda^2(s'-s),\quad s<s'\in ]0;\Delta].
\end{equation}
Eqs. (\ref{rozpds0}) and (\ref{rozpdr0})($*$) describe situation when one ISI starts with impuls in the feedback line, which has time to live equal $s$, and the next ISI starts with impuls in the line, which has time to live equal $\Delta$. Thus, $P(s\mid s')$ has singularity of $\delta$-function type at $s=\Delta$. For calculating its mass, one should take (\ref{rozpds0}), (\ref{rozpdr0})($*$) with $s$ replaced with $s'$ and calculate integral over admittable values of $t$:
$$
e^{-\lambda s'}\lambda s'
+\int\limits_{s'}^{s'+\tau}e^{-\lambda t}\lambda\,dt+
\int\limits_{s'+\tau}^\infty 
e^{-\lambda (\tau+s')}
P^0(t-s'-\tau)\,dt=
\lambda\,s'\, e^ {- \lambda\,s' }+e^ {- \lambda\,s' }.
$$
Here we use $\int\limits_0^\infty\,P^0(t)\,dt=1$. 
Thus, $P(s\mid s')$ is the sum of two functions
\begin{equation}\label{umov2}
P(s\mid s')=P_1(s,s')+P_2(s,s'),
\end{equation}
where
\begin{align*} 
P_1(s,s')&=
\begin{cases}
e^{-\lambda(s'-s)}\lambda^2(s'-s)\text{ при } s<s'\in ]0;\Delta]\\
0\text{ при } s\ge s'
\end{cases},
\\
P_2(s,s')&=\delta(s-\Delta)\left(\lambda\,s'\, e^ {- \lambda\,s' }+e^ {- \lambda\,s' }\right).
\end{align*}
The transition probability $P(s\mid s')$ is normed:
$
\int\limits_0^\Delta P(s\mid s')\,ds=1.
$

\subsection{Delays distribution}

Here we found probability density distribution $f(s)$. For this purpose let us represent $f(s)$ as
\begin{equation}\label{vygliad}
f(s)=a\,\delta(s-\Delta)+g(s)=a\,\delta(s-\Delta)+e^{\lambda s}\varphi(s),
\end{equation}
where $a$ --- is a dimensionless constant, and $g(s), \varphi(s)$ --- are ordinary functions. After substituting (\ref{umov2}) and (\ref{vygliad}) into Eq. (\ref{umoriv}), and separating terms without $\delta$-function, one obtains
$$
a\,e^{-\lambda\Delta}\lambda^2(\Delta-s)
+\lambda^2\int\limits_s^\Delta (s'-s)\varphi(s')\,ds'=\varphi(s).
$$
This equation can be easily solved with respect to $\varphi(s)$, which delivers $g(s)$ as
\begin{equation}\label{groz}
g(s)=\frac{a\,\lambda}{2}\left(1-e^{- 2\lambda(\Delta-s)}\right).
\end{equation}
Now take into account that $f(s)$ must be normed:
$
a+\int\limits_0^\Delta g(s)\,ds=1,
$
 which gives for $a$:
\begin{equation}\label{a}
a=\frac{4e^{2\lambda\Delta}}{(2\lambda\Delta+3)e^{2\lambda\Delta}+1}.
\end{equation}

\subsection{ISIs distribution}

\begin{figure}
\begin{center}
\includegraphics[width=0.66\textwidth,angle=0]{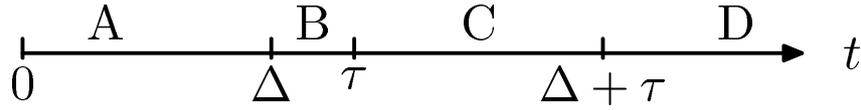}
\end{center}
\caption{\label{t-cases0}%
Domains of $t$ used for calculating integral in (\ref{Pu0}).}
\end{figure}

For calculating $P^\Delta(t)$ substitute (\ref{rozpsum}), (\ref{rozpds0}), (\ref{rozpdr0}) and 
(\ref{vygliad}), (\ref{groz}), (\ref{a}) into Eq. (\ref{usered}). This gives
\begin{equation}\label{Pu0}
P^\Delta(t)=e^{-\lambda t}\lambda t (a \delta(t-\Delta) + g(t)) +
a P^{\Delta r}_\Delta(t) +
\int\limits_0^\Delta P^{\Delta r}_s(t)\,g(s)\,ds.
\end{equation}
Further transformation of (\ref{Pu0}) depends on the $t$ value. Basic domais of $t$ are shown in Fig. \ref{t-cases0}.

Consider case A. Here integration domain should be splitted into two with point $s=t$. This gives
$$
P^\Delta(t)=e^{-\lambda t}\lambda t  g(t) +
a \lambda^2 t e^{-\lambda t} +
\int\limits_0^t \lambda e^{-\lambda t}\,g(s)\,ds+
\int\limits_t^\Delta \lambda^2 t e^{-\lambda t}\,g(s)\,ds,
$$
which after transformations becomes
\begin{equation}\label{t-A}
P^\Delta(t)={e^{-\lambda\,t}\left((2\,\lambda\Delta +7) \lambda t\, e^{2\lambda\Delta}+1-(\lambda t+1)e^{2\lambda t}-2\, \lambda^2\,t^2\,e^{2\,\lambda\Delta }\right)\lambda\over{
 \left(2\,\lambda\Delta +3\right)\,e^{2\,\lambda\Delta }+1}},\quad t<\Delta.
\end{equation}

At $t=\Delta$, ISI distribution $P^\Delta(t)$ has $\delta$-function type singularity:
\begin{equation}\label{t-Delta}
P^\Delta(t)={4\, \lambda \,\Delta\,e^{\lambda\Delta}\over\left(2\,\Delta\,
  \lambda +3\right)\,e^{2\,\lambda\Delta }+1}\,
\delta(t-\Delta),\quad t\in]\Delta-\epsilon;\Delta+\epsilon[.
\end{equation}

Consider case B. Here integration in (\ref{Pu0}) can be performed over the entire domain $]0;\Delta[$ uniformly, which gives
\begin{equation}\label{t-B}
P^\Delta(t)=e^{-\lambda t}\lambda\,\int\limits_0^\Delta f(s)\,ds=
e^{-\lambda t}\lambda,\quad \Delta<t<\tau.
\end{equation}

Consider case C. Here integration domain should be splitted into two with point $s=t-\tau$, and Eq. (\ref{Pu0}) turns into the following:
$$
P^\Delta(t)=
\int\limits_0^{t-\tau} e^{-\lambda (\tau+s)}
P^0(t-s-\tau) g(s)\,ds+
e^{-\lambda t}\lambda\int\limits_{t-\tau}^\Delta  g(s)\,ds
+a\,e^{-\lambda t}\lambda.
$$
Here in the first integral $(t-s-\tau)\in[0;t-\tau]\subset[0;\Delta]\subset[0;\tau]$. This allows 
to identify from Eq. (\ref{P0}) exact expression for $P^0(t-s-\tau)$, which is  $e^{-\lambda(t-s-\tau)}\lambda^2(t-s-\tau)$:
$$
P^\Delta(t)=
\int\limits_0^{t-\tau} 
e^{-\lambda t}\lambda^2(t-s-\tau) g(s)\,ds+
e^{-\lambda t}\lambda\int\limits_{t-\tau}^\Delta  g(s)\,ds
+a\,e^{-\lambda t}\lambda.
$$
After transformations, one obtains
\def\kt{e^{\lambda t}}
\def\ktta{e^{-2\lambda (t-\tau)}}
\def\kd{e^{2\lambda \Delta}}
\def\denom{\kd (4\lambda \Delta + 6) + 2}
\def\la{\lambda}
\def\ta{\tau}
\begin{equation}\label{t-C}
P^\Delta(t)=
\frac{(K_0+K_1 t+K_2 t^2+e^{2\la(t-\ta)})\la e^{-\la t}}{\denom},\quad
\tau<t<\Delta+\tau,
\end{equation}
where
\begin{align*}
K_0&=\left(2\lambda^2\tau^2+4\lambda\tau+4\lambda\Delta+6\right)e^{2\la\Delta}-2\la\ta+1,
\\
K_1&=
(2-4e^{2\,\lambda\,\Delta}(1+\lambda\,\tau))\lambda,\quad
K_2=
2\,\lambda^2\,e^{2\,\lambda\,\Delta}.
\end{align*}

Consider case D. Here Eq. (\ref{Pu0}) turns into the following:
$$
P^\Delta(t)=
\int\limits_0^\Delta e^{-\lambda (\tau+s)}
P^0(t-s-\tau) g(s)\,ds+a\,e^{-\lambda (\tau+\Delta)}P^0(t-\Delta-\tau).
$$
Let us introduce a new variable of integration, $u=t-s-\tau$:
\begin{equation}\label{t-Du}
P^\Delta(t)=
\int\limits_{t-\Delta-\tau}^{t-\tau} e^{-\lambda (t-u)}
P^0(u) g(t-\tau-u)\,du+a\,e^{-\lambda (\tau+\Delta)}P^0(t-\Delta-\tau),
\end{equation}
From this expression we see, that for calculating the integral one needs to use Eq. (\ref{P0}) either with single, or with two consecutive values of $m$. Namely, if for some $m$:\, $m\tau\le t-\Delta-\tau<t-\tau\le(m+1)\tau$, then one should substitute term from (\ref{P0}), corresponding to that $m$ instead of $P^0(u)$ in the (\ref{t-Du}). In the opposite situation, there exist such $m$, that $m\tau< t-\Delta-\tau<(m+1)\tau<t-\tau$. In this case, domain of integration in the Eq. (\ref{t-Du}) should be split with point $(m+1)\tau$, and as $P^0(u)$ one should substitute term from (\ref{P0}), corresponding either to $m$, or to $m+1$. Thus, when $t\in [\Delta+\tau;\infty[$, then all possible situations are parameterized with the above mentioned number $m$ in such a way that if $t\in[\Delta+(m+1)\tau;(m+2)\tau[$, then use term from (\ref{P0}) with that $m$, and if $t\in[(m+2)\tau;\Delta+(m+2)\tau[$, then split integration domain and use terms with both $m$, and $m+1$.

\begin{figure}
\begin{center}
\includegraphics[width=0.33\textwidth,angle=-90]{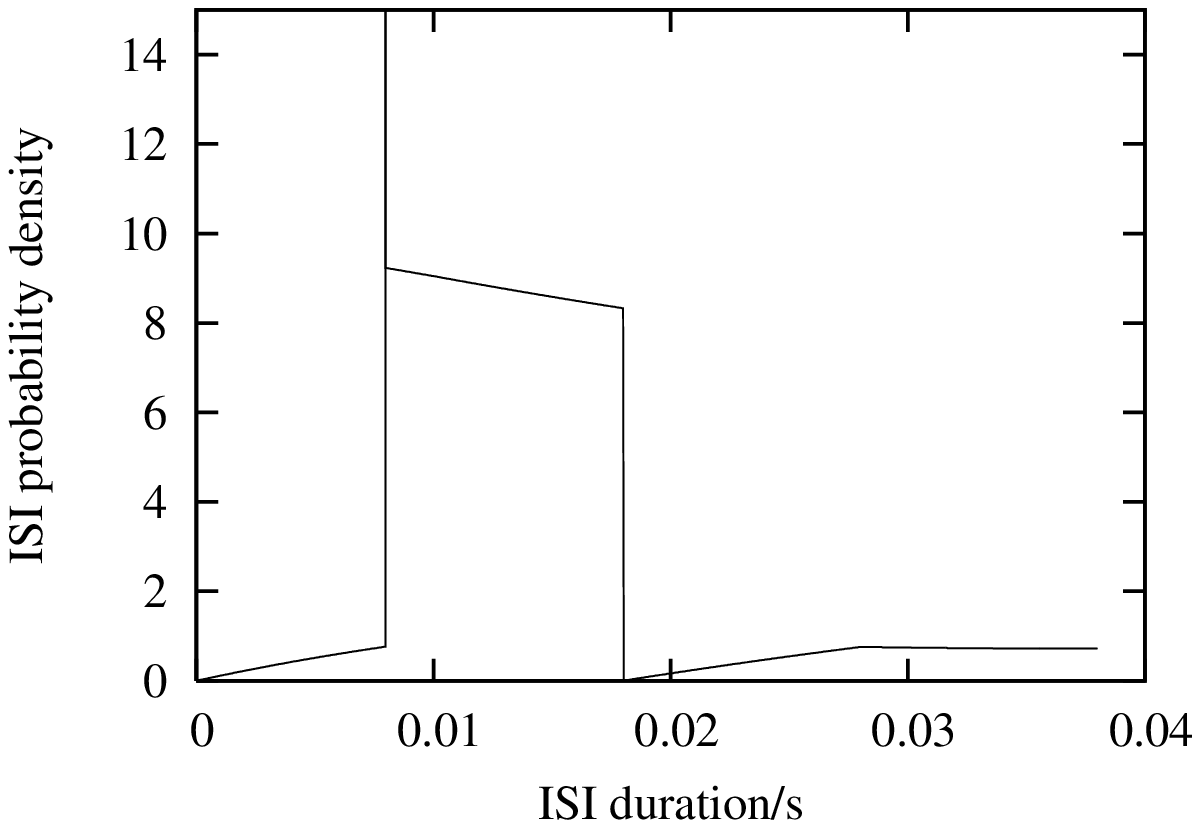}
\includegraphics[width=0.33\textwidth,angle=-90]{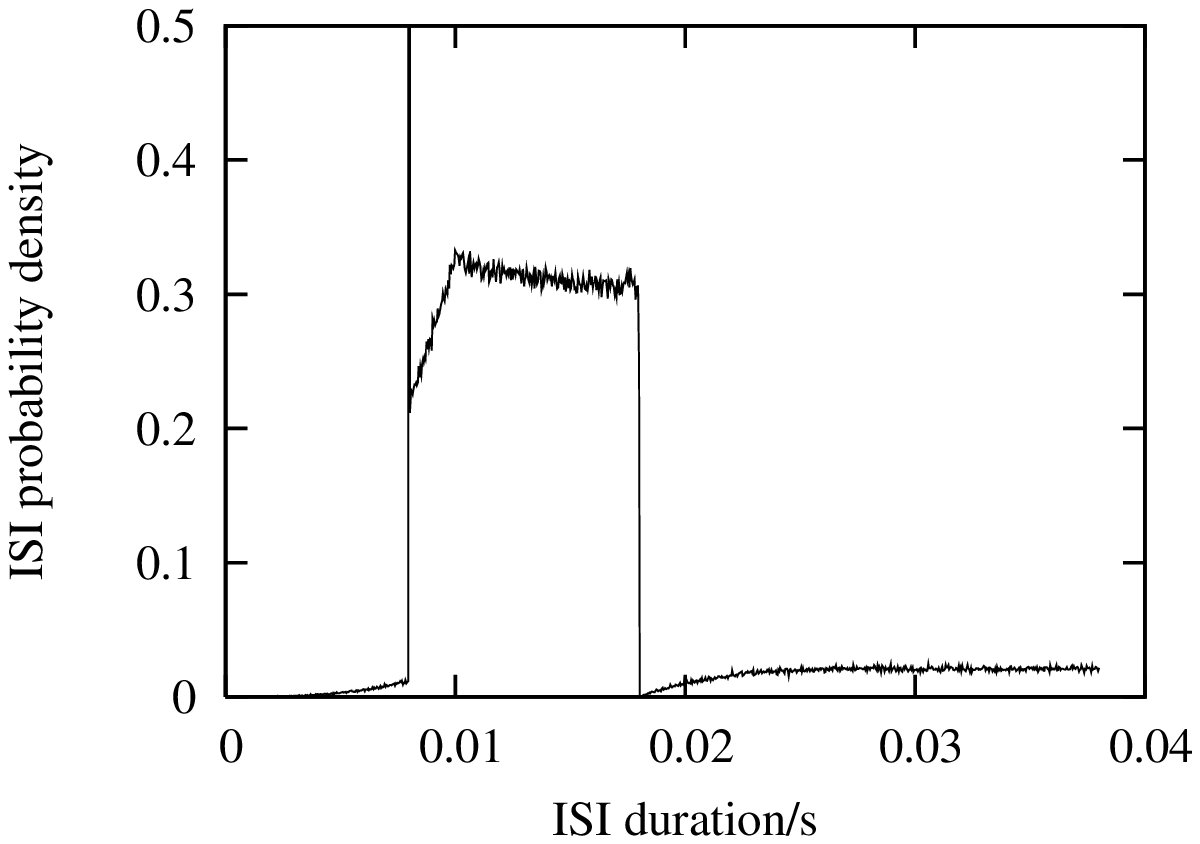}
\end{center}
\caption{\label{ISI}%
Example of ISI probability density function, calculated in accordance with Eqs. (\ref{t-A}),(\ref{t-Delta}),(\ref{t-B}),(\ref{t-C}),(\ref{t-D1}), left panel, and numerically, by means of Monte Carlo method, right panel. For both panels:  $\tau=10$ ms, $\Delta=8$ ms, $\lambda=10$ s$^{-1}$. In the left panel, $N_0=2$, in the right, $N_0=4$. Curve found numerically for $N_0=2$ fits perfectly with one shown in the left panel. In the numerical experiment 360000000 spikes were produced.}
\end{figure}

For example, if $t\in[\Delta+\tau;2\tau[$, then $m=0$ and (\ref{t-Du}) turns into
$$
P^\Delta(t)=
\int\limits_{t-\Delta-\tau}^{t-\tau} e^{-\lambda (t-u)}
e^{-\lambda u}\lambda^2 u g(t-\tau-u)\,du+a\,e^{-\lambda t}\lambda^2(t-\Delta-\tau),
$$
which gives after transformations
\begin{equation}\label{t-D1}
P^\Delta(t)=\lambda^2(t-\tau) e^{-\lambda t} +\frac{1-\left(2\,\Delta^2\,\lambda^2+6\,\Delta\,\lambda+1\right)\,\kd}{\denom}\,\lambda e^{-\lambda t},\quad t\in[\Delta+\tau;2\tau[.
\end{equation}
Graph of $P^\Delta(t)$ is shown in Fig. \ref{ISI}.

\subsection{Output intensity}

Let us found mean output ISI, $W^\Delta$. Output intensity is inversed $W^\Delta$. The $W^\Delta$ is defined as
$$
W^\Delta = \int\limits_0^\infty t P^\Delta(t)\,dt.
$$
Use here Eq.  (\ref{usered}):
$$
W^\Delta = \int\limits_0^\infty t \,dt\,\int\limits_0^\Delta P_s^d(t)f(s)\,ds
=
\int\limits_0^\Delta ds\,f(s)\int\limits_0^\infty t P_s^d(t)\,dt.
$$
Use here representation (\ref{rozpds0}), (\ref{rozpdr0}) and Eq. (\ref{W1}):
\begin{multline*}
W^\Delta  =
\int\limits_0^\Delta ds\,f(s)
\left(
\int\limits_0^s t^2 e^{-\lambda t}\lambda^2\,dt +
e^{-\lambda s}\lambda s^2 +
\int\limits_s^{s+\tau} t \lambda e^{-\lambda t}\,dt\right)
+
\\
+
\int\limits_0^\Delta ds\,f(s)\, e^{-\lambda(\tau+s)}
\int\limits_{s+\tau}^\infty t  P^0(t-s-\tau)\,dt
=\\=
\int\limits_0^\Delta ds\,f(s)
\frac{2-(1+\lambda s)e^{-\lambda s} -(1+\lambda\tau+\lambda s)e^{-\lambda(\tau+s)}}{\lambda}+
\\
+\int\limits_0^\Delta ds\,f(s)\, e^{-\lambda(\tau+s)}
\left(
s+\tau + \frac{1}{\lambda}\left(2+\frac{1}{e^{\lambda\tau}-1}\right)
\right)\,.
\end{multline*}
Use here (\ref{vygliad}), (\ref{groz}), (\ref{a}), which gives after transformations:
\begin{equation}\label{Wd}
W^\Delta=\frac
{
2\left(\left(2\lambda\Delta+e^{-2\lambda\Delta}+1\right)-2\lambda\Delta e^{-\lambda\tau}\right)
}{
\lambda\left(2\lambda\Delta+e^{-2\lambda\Delta}+3\right)\,\left(1-e^{-\lambda\tau}\right)
}.
\end{equation}
The output intensity is then
$
\lambda^\Delta_o = 1/W^\Delta.
$
At large input rates the following relation takes place
\begin{equation}\label{lala}
\lim\limits_{\lambda\to\infty} \left(\lambda^\Delta_o -\frac{\lambda}{2}\right)=\frac{1}{2\Delta}.
\end{equation}

\section{Numerical simulation}

In order to check correctness of obtained analytical expressions, as well as to obtain an impression how ISI distribution looks like for higher thresholds, a C++ program was developed, which directly modells functioning of BN with delayed feedback. The Poisson input streams were generated by transformation of uniformly distributed sequences of random numbers (see, e.g. Eq. (12.14) in  \cite{Cell}). The ISI probability density is found by counting output ISI of different durations and normalization. In the program, distribution of time to live of impulse in the feedback line was calculated as well. Numerically obtained curves fit perfectly with the analytical expressions for $P^\Delta(t)$ given in Eqs. (\ref{t-A}),(\ref{t-Delta}),(\ref{t-B}),(\ref{t-C}),(\ref{t-D1}), and for $f(s)$ given in Eqs. (\ref{vygliad}), (\ref{groz}), (\ref{a}).

\section{Conclusions and discussion}

We calculatet here ISI probability density functions for binding neuron with delayed feedback. For BN with threshold 2 ISI distribution is found analytically and numerically, and for threshold 4 --- numerically. The function obtained have remarkable peculiarities which suggests what could happen with spiking statistics of individual neurons in elaborated network with delayed connections. For threshold 2 we also found the output intensity as a function of the input one. The limiting relation (\ref{lala}) can be understood as follows. At moderate stimulation some input spikes are lost without influencing output due to high probability of long input ISI. At high intensity every two consecutive input impulses trigger the BN and send impuls into the feedback line, provided it is empty. Thus, output intensity should be $\lambda/2$ plus firing, caused by additional stimulation from the line. This additional stimulation has maximum rate $1/\Delta$, which explaines (\ref{lala}).


\end{document}